\documentclass[epjST]{svjour}
\usepackage{graphicx}
\usepackage{amsmath}
\usepackage{amssymb}
\usepackage[greek,english]{babel}
\usepackage{teubner}
\usepackage[numbers,square,comma,sort&compress]{natbib}
\begin{document}
\bibliographystyle{prsty}
\def\qq{{\hbox{\foreignlanguage{greek}{\coppa}}}}
\def\qqq{{\hbox{\foreignlanguage{greek}{\footnotesize\coppa}}}}
\graphicspath {{./figures/}}
\makeatletter
\def\input@path{{./figures/}}
\makeatother

\title{Cluster Monte Carlo and dynamical scaling for long-range interactions}

\author{
  E. Flores-Sola\inst{1,2,3}
  \and M. Weigel\inst{1,3}\fnmsep\thanks{\email{martin.weigel@coventry.ac.uk}}
  \and R. Kenna\inst{1,3}
  \and B. Berche\inst{2,3}
}
\institute{
  Applied Mathematics Research Centre, Coventry University, Coventry CV1 5FB, United Kingdom
  \and Institut Jean Lamour, CNRS/UMR 7198, Groupe de Physique Statistique,
  Universit\'e de Lorraine, BP 70239, F-54506 Vand\oe uvre-les-Nancy Cedex, France
  \and {Doctoral College for the Statistical Physics of Complex Systems,
    Leipzig-Lorraine-Lviv-Coventry $({\mathbb L}^4)$}
}

\abstract{
  Many spin systems affected by critical slowing down can be efficiently simulated
  using cluster algorithms. Where such systems have long-range interactions, suitable
  formulations can additionally bring down the computational effort for each update
  from O($N^2$) to O($N\ln N$) or even O($N$), thus promising an even more dramatic
  computational speed-up. Here, we review the available algorithms and propose a new
  and particularly efficient single-cluster variant. The efficiency and dynamical
  scaling of the available algorithms are investigated for the Ising model with
  power-law decaying interactions.
}

\maketitle

\section{Introduction}
\label{secIntroduction}

The theory of phase transitions and critical phenomena is by now rather well
understood, although there remain a significant number of questions that are still
very actively debated and some of which are not finally settled, ranging from the
theory of disordered systems \cite{kawashima:03a,fytas:16} to quite fundamental
problems such as certain aspects of finite-size scaling
\cite{wittmann:14,flores:16}. The theoretical basis of this success is the concept of
the renormalization group \cite{fisher:74a} that allows one to understand scaling and
universality in such systems. While this theory provides the essential scaffolding
for describing continuous phase transitions, and many results have been derived from
it via perturbative approaches such as the $\epsilon$ expansion, the field is now
hardly conceivable without contributions from numerical techniques such as the
molecular dynamics \cite{rapaport:04} and Monte Carlo methods
\cite{binder:book2}. While the basic techniques such as, e.g., the Metropolis
algorithm \cite{metropolis:53a} for simulations of spin systems, are rather easily
implemented, using them for studies of critical points is not straightforward as
there the simulational and the physical dynamics are affected by critical slowing
down due to the proliferation of spatial correlations as the critical point is
approached. An effective antidote for this problem is available in the form of {\em
  cluster algorithms\/} as originally proposed for the short-range Ising model
\cite{swendsen-wang:87a,wolff:89a} and later generalized to a range of different spin
models and certain off-lattice systems \cite{luijten:06}. They manage to
substantially reduce or, in some cases, practically eliminate critical slowing down
through the identification and updating of large-scale, fractal structures whose
extent diverges as the critical point is approached. A related class of algorithms
even achieve critical speeding up, an increase of computational efficiency with
system size, for certain quantities \cite{deng:07,elci:13}.

While the electromagnetic force as the basic agent in condensed-matter systems decays
slowly, proportional to the inverse square of the distance, due to screening effects
short-range intermolecular interactions such as those parameterized in the
Lennard-Jones potential dominate in many cases. In some systems, however, such as in
frustrated magnets \cite{lacroix:11} or in certain lattices of cold atoms
\cite{britton:12}, long-range interactions are responsible for the presence or
absence of ordering. The effect of such interactions on the nature of the transition
was studied early on in the framework of the renormalization group
\cite{fisher:72}. Apart from describing experimentally relevant systems with
long-range interactions, these models were also soon recognized as pathways for
introducing non-trivial critical behavior into systems whose physical dimension is
too low to exhibit such ordering for short-range couplings
\cite{dyson:69,dyson:71}. While both local and cluster-update Monte Carlo simulation
algorithms are directly applicable to systems with long-range interactions, the fact
that each particle or spin interacts with all others implies an $O(N^2)$ scaling of
the computational effort for the simulation of a system of $N$ particles. As a
result, the system sizes accessible computationally through these methods are
severely restricted, typically to a few thousand spins \cite{xu:93}. Due to this
limitation, many studies considered cut-offs to the interactions and/or extrapolation
methods to try to access ranges of system sizes where finite-size scaling approaches
could be meaningfully employed \cite{nagle:70,glumac:89}. It was realized by Luijten
and Bl\"ote \cite{luijten:95} that these restrictions could be lifted using a
different formulation of the cluster algorithm that allows one to update all spins
once with an $O(N\log N)$ computational effort. More recently, Fukui and Todo
\cite{fukui:09} proposed a related, slightly more general approach with a scaling
only slightly worse than linear. Below, we discuss a single-cluster variant with
strictly $O(N)$ scaling. These methods hence deliver a twofold and very dramatic
speedup: a computational acceleration from $O(N^2)$ to $O(N)$ operations per sweep
and, in addition, a reduction of critical slowing down in the vicinity of the
critical point.

The rest of the paper is organized as follows. In Sec.~\ref{sec:SW} we give a short
summary of the Swendsen-Wang algorithm to set the stage for the improved
methods. Sections \ref{sec:LB} and \ref{sec:FT} discuss the Luijten-Bl\"ote and
Fukui-Todo approaches, respectively, including the single-cluster variant introduced
here. The computational and algorithmic performance is discussed in
Sec.~\ref{sec:scaling}. Finally, Sec.~\ref{sec:conclusions} contains our conclusions.

\section{Swendsen-Wang algorithm}
\label{sec:SW}

Although the algorithms discussed below can be generalized to the case of Potts and
even continuous-spin models, for the sake of simplicity we restrict our presentation
to the case of the Ising model with Hamiltonian
\begin{equation}
{\cal H} = -\sum_{i,j} J_{ij} s_i s_j - \sum_i H_i s_i,\;\;\;s_i = \pm 1.
\label{eq:hamiltonian}
\end{equation}
Here, the sum is over all lattice sites, $J_{ij}$ is the exchange coupling between
spins $i$ and $j$ and $H_i$ denotes a local external magnetic field. Unless stated
otherwise, we will focus on the case of zero external fields, $H_i = 0$.

Let us first consider the case of homogeneous nearest-neighbor interactions,
$J_{ij} = J$ if $i$, $j$ are nearest neighbors on the lattice and $J_{ij} = 0$
otherwise. It is straightforward to verify the following identities for the partition
function of the model \cite{wj:chem},
\begin{equation}
  \begin{split}
    Z &= \sum_{\{s_i\}} \exp\left(\beta J \sum_{\langle i,j\rangle} s_i s_j\right) \\
    &= \sum_{\{s_i\}} \prod_{\langle i,j\rangle} e^{\beta J}
    \left[(1-p)+p\delta_{s_i,s_j}\right] \\
    &=  \sum_{\{s_i\}} \sum_{n_{ij}} \prod_{\langle i,j\rangle}
    e^{\beta J}
    \left[(1-p)\delta_{n_{ij},0}+p\delta_{s_i,s_j}\delta_{n_{ij},1}\right] \\
    &= \sum_{\{s_i\}} \sum_{n_{ij}} \prod_{\langle i,j\rangle} W(s_i,s_j,n_{ij}).
  \end{split}
  \label{eq:RCM}
\end{equation}
Here, $n_{ij} \in \{0,1\}$ are new, binary variables that represent the state of the
bonds as `active' or `deleted', and $p = 1-\exp(-2\beta J)$ is the bond activation
probability. The form (\ref{eq:RCM}) corresponds to the Fortuin-Kasteleyn (FK)
representation of the Ising model \cite{fortuin:72a}. This transformation from a pure
spin model to a probability measure jointly defined on spins and `graph' variables
(i.e., bonds), known as the Edwards-Sokal coupling \cite{edwards:88a}, is at the
heart of all cluster updates of this type. The representation (\ref{eq:RCM}) implies
the following update procedure known as the Swendsen-Wang algorithm
\cite{swendsen-wang:87a} for the ferromagnetic, nearest-neighbor Ising model:
\begin{enumerate}
\item Activate bond variables, i.e., set $n_{ij} = 1$, between neighboring spins with
  probability $P_{ij} = \delta_{s_i,s_j}p$.
\item Identify clusters of spins connected by active bonds.
\item Flip independent clusters with probability $1/2$.
\end{enumerate}
This generates a new spin configuration which is then subjected to a new iteration of
the same procedure etc. As it is possible to have single-site clusters, the algorithm
is ergodic. Together with detailed balance, which can be shown quite
straightforwardly by inspecting the configuration weight $W(s_i,s_j,n_{ij})$ in the
joint spin and bond space, this guarantees that the underlying Markov chain converges
to the equilibrium stationary distribution \cite{binder:book2}. For the best possible
choices of the algorithm used for cluster identification \cite{weigel:10b}, one full
update of the Swendsen-Wang algorithm requires $O(N+E)$ operations, where $N$ is the
number of spins and $E$ is the number of edges in the graph. For a short-range
lattice model, $E=zN/2$, where $z$ is the coordination number, resulting in $O(N)$
scaling.

Consider now a long-range model where, in general, all $J_{ij}$ are non-zero and
different. The FK representation (\ref{eq:RCM}) is easily generalized to this case by
noting that the weight function is now
\[
 W(s_i,s_j,n_{ij}) =  e^{\beta J_{ij}}
    \left[(1-p_{ij})\delta_{n_{ij},0}+p_{ij}\delta_{s_i,s_j}\delta_{n_{ij},1}\right],
\]
where the bond dependent activation probability is
\[
p_{ij} = 1-\exp(-2\beta J_{ij}).
\]
Note that there is a subtlety in the notation here, with
$P_{ij} = \delta_{s_i,s_j}p_{ij}$ being the activation probability for the bond
between $i$ and $j$, while $p_{ij}$ is the activation probability {\em conditioned\/}
on $s_i = s_j$. The resulting cluster algorithm still satisfies ergodicity and
detailed balance, so is correct. As $E = N^2-N$, however, one update is now much more
expensive, and we expect $O(N^2)$ run-time scaling asymptotically, much like that for
single spin updates.

A variant of the Swendsen-Wang algorithm due to Wolff \cite{wolff:89a} grows and
flips only a single cluster, emanating from a single, randomly chosen seed site, thus
reducing the effort for cluster identification (which can be done on the fly) and
leads to larger clusters being flipped, such that it is typically more
efficient. While run times are now proportional to the number of edges in a given
cluster, the number of operations required to update each spin once on average
remains $O(N^2)$.

\section{Luijten-Bl\"ote algorithm}
\label{sec:LB}

Thus, while the Swendsen-Wang or Wolff (single cluster) algorithms naturally extend
to the case of long-range interactions, their direct application leads to $O(N^2)$
scaling. Luijten and Bl\"ote \cite{luijten:95} noticed that in the interesting regime
of couplings and temperatures most probabilities $p_{ij}$ will be very small and so
there are many rejected bond activation attempts. These can be avoided by directly
sampling from the cumulative distribution of bond probabilities. To see this,
consider for definiteness a system with power-law interactions,
\begin{equation}
  J_{ij} = \frac{J}{r_{ij}^{d+\sigma}}.
  \label{eq:potential}
\end{equation}
The bond activation probabilities are then
$p_{ij} = 1-\exp(-2\beta J/r_{ij}^{d+\sigma}) \equiv p_r$, and only depend on the
distance $r=r_{ij}$ of lattice sites.  In the following, we only make use of this
translational invariance and not of the specific power-law form of
Eq.~(\ref{eq:potential}). We first consider the special case of a chain, i.e.,
lattice dimension $d=1$.  The algorithm is a single-cluster variant, where spins are
added successively to the cluster starting from an initial, random seed site by
probing the bonds emanating from the currently considered spin for activation. If the
current spin is at site $i$, we consider spins at sites $j$ to be added in the order
of increasing distance $|j-i|$ along the chain. The probability that the first bond
connecting to $i$ to be activated in this way is the spin at site $k$ is
\[
q(k) = (1-p_1)(1-p_2)\cdots(1-p_{k-1})p_k.
\]
We can pick a site $k$ according to this probability by considering the cumulative
distribution,
\begin{equation}
  C(k) = \sum_{n=1}^k q(n) = 1-\prod_{n=1}^{k}(1-p_n).
  \label{eq:cumu1}
\end{equation}
If a random number $r$ drawn uniformly in $[0,1[$ is found to be between $C(k-1)$ and
$C(k)$, the next spin to be added to the cluster is at distance $k$. The next bond
after that must be drawn between the current spin $i$ and another spin at distance
$l>k$, and so the relevant probability is
\[
q_k(l) = (1-p_{k+1})(1-p_{k+2})\cdots(1-p_{l-1})p_l,
\]
and we need to sample from the cumulative distribution
\begin{equation}
  C_k(l) = \sum_{n=k+1}^l q_k(n) = 1-\prod_{n=k+1}^{l}(1-p_n),
  \label{eq:cumu2}
\end{equation}
such that $C(k) = C_0(k)$. For the third and higher spins we proceed iteratively
along the same lines.

Note that for free boundaries we also need to allow for the possibility of activating
bonds to spins with $k < i$, i.e., to the left of the current spin. This can be taken
into account by formally using an interaction strength $2J$ instead of $J$ and using
an extra random number for each activated bond to decide whether it is to a spin to
the left or to the right of the current one. For periodic boundaries, on the other
hand, the distance definition to use (for the chain) is modified to
$r_{ij} = \min(|i-j|,L-|i-j|)$.

This leaves us to decide how efficiently one can sample from the cumulative
distributions (\ref{eq:cumu1}) and (\ref{eq:cumu2}), respectively.  We note that
$C(k)$ and $C_k(l)$ are related by a linear transformation,
\[
C_k(l) = \frac{p_{k+1}}{q(k+1)} \left[C(l)-C(k)\right],
\]
such that it is sufficient to store $N$ elements in such a look-up table for a
translationally invariant system. In the most naive implementation, we would require
a number of comparisons that is $O(N)$ to decide in which bin the random number $r$
falls. This can be of course be avoided using a binary search in a table of the
probabilities $C_k(l)$, reducing the effort to a factor $\log N$. If one wants to
truncate the interaction to reduce the storage and time effort for look-ups, this
works less well in higher dimensions as the number of distinct lattice distances to
store in the vicinity of the seed site grows quickly with $d$. Luijten and Bl\"ote
hence suggest \cite{luijten:95,luijten:97} to modify the interaction potential in a
way that allows for an analytic inversion of the cumulative distribution function and
to thus avoid the lookup tables altogether. This will not lead to correct results for
the original model considered, but it can be an affordable simplification if one is
only interested in universal quantities which are independent of such details. A
direct generalization of the exact algorithm to higher dimensions is feasible, but a
little bit tedious due to the necessary bookkeeping.

The approach of Luijten and Bl\"ote thus entails a computational effort of
$O(\log E) = O(\log N)$ for the look-up of each weight (even in the
non-translationally invariant case), and since there are on average $O(N)$ active
bonds in the regime where the potential (\ref{eq:potential}) is integrable
\cite{fukui:09}, the total effort is $O(N\log N)$.

\section{Fukui-Todo algorithm}
\label{sec:FT}

The cluster update could be simplified further if one could decide about the number
of active bonds at the onset and place them according to the local bond weights. This
is in fact possible if one allows the bond activation variables that are restricted
to $n_{ij} \in \{0,1\}$ following the FK representation to take arbitrary,
non-negative integer values $n_{i,j} = 0$, $1$, $2$, $\ldots$ \cite{fukui:09}. This
is compatible with the FK weight if one ensures that the probability of a non-zero
$n_{ij}$ for parallel spins is identical to the standard bond activation probability
$p_{ij}$, i.e.,
\begin{equation}
  \mathbb{P}(n_{ij} > 0 | s_i = s_j) = \sum_{n_{ij}=1}^\infty f(n_{ij}) = p_{ij}
  = 1-e^{-2\beta J_{ij}}.
  \label{eq:norm}
\end{equation}
Instead of the binary distribution
$f(n_{ij}) = (1-p_{ij})\delta_{n_{ij},0} + p_{ij}\delta_{n_{ij},1}$ of the standard
FK model, we now choose a Poisson distribution,
\begin{equation}
  \label{eq:poisson}
  f_\lambda(n) = \frac{e^{-\lambda}\lambda^n}{n!},
\end{equation}
where the normalization condition (\ref{eq:norm}) implies that
$\lambda_{ij} = 2\beta J_{ij}$. This can be formally incorporated into a generalized
FK representation by noting that the FK weight can be factorized into the $p_{ij}$
dependent part that only contains $n_{ij}$, while the remainder depending on $s_is_j$
is independent of $p_{ij}$,
\[
W(s_i,s_j,n_{ij}) = e^{\beta J_{ij}} V(n_{ij})\Delta(s_i,s_j,n_{ij}),
\]
where
\begin{equation}
  \begin{split}
    V(n_{ij}) &=  (1-p_{ij})\delta_{n_{ij},0} + p_{ij}\delta_{n_{ij},1}, \\
    \Delta(s_i,s_j,n_{ij}) &= \delta_{n_{ij},0} + \delta_{s_i,s_j}\delta_{n_{ij},1}.
  \end{split}
\end{equation}
We can therefore write down the FK representation of the model with general integer
bond variables, 
\begin{equation}
  \begin{split}
    V(n_{ij}) &=  \frac{e^{-2\beta J_{ij}} (2\beta J_{ij})^{n_{ij}}}{n_{ij}!}, \\
    \Delta(s_i,s_j,n_{ij}) &= \delta_{n_{ij},0} + (1-\delta_{n_{ij},0})\delta_{s_i,s_j}.
  \end{split}
  \label{eq:RCMgen}
\end{equation}

The advantage of allowing the bond variables to take on arbitrary integer values
$n_{ij} \ge 0$ according to a Poisson distribution is that any (finite or infinite)
sum of Poisson random variables, even of different means, leads again to a Poisson
distribution, i.e., it is a sum-stable distribution. Additionally, we have the
following distribution identity,
\begin{equation}
  \prod_{(i,j)} f_{\lambda_{ij}}(n_{ij}) = f_{\lambda_\mathrm{tot}}(n_\mathrm{tot})
  \frac{n_\mathrm{tot}!}{\prod_{(i,j)}n_{ij}!}
  \prod_{(i,j)} \left(\frac{\lambda_{ij}}{\lambda_\mathrm{tot}}\right)^{n_{ij}},
  \label{eq:composition}
\end{equation}
where $(i,j)$ denotes the (undirected) bond connecting the sites $i$ and $j$ on the
lattice.  The left-hand side corresponds to drawing each $n_{ij}$ independently
according to $f_{\lambda_{ij}}$. The right-hand side, on the other hand, represents a
prescription where a total number $n_\mathrm{tot}$ is drawn according to
$f_{\lambda_\mathrm{tot}}$ with
$\lambda_\mathrm{tot} = \sum_{(i,j)} \lambda_{ij} = \sum_{(i,j)} 2\beta J_{ij} =
2\beta J_\mathrm{tot}$
first and these `events' are then randomly distributed over the actual bonds with a
probability $\lambda_{ij}/\lambda_\mathrm{tot}$. The equation expresses the fact that
these two procedures lead to the same final distribution of $n_{ij}$. The
distribution of events can be performed using tables with binary search as for the
approach of Luijten and Bl\"ote or, alternatively, using Walker's method of alias, as
will be discussed below in Sec.~\ref{sec:FTmulti}.

\subsection{Multi-cluster method}
\label{sec:FTmulti}

The modified FK representation can be used to simulate the underlying Ising model as
follows:

\begin{enumerate}
\item Draw a total number of events $n_\mathrm{tot} \ge 0$ according to a Poisson
  distribution with mean $\lambda_\mathrm{tot} = 2\beta \sum_{(ij)} J_{ij}$.
\item Distribute each event to one of the bonds with probability
  $\lambda_{ij}/\lambda_\mathrm{tot}$.
\item Identify clusters of like spins connected by bonds with $n_{ij} > 0$ and flip
  each cluster with probability $1/2$.
\end{enumerate}

The normalization (\ref{eq:norm}) ensures that the actual spin dynamics of this
approach is the same as that of the Swendsen-Wang approach. It is different, however,
from the Luijten-Bl\"ote algorithm in the same sense as the Wolff algorithm is
different from Swendsen-Wang dynamics since in the single-cluster algorithm on
average larger clusters are flipped.

Let us discuss some implementation details and the computational complexity of each
step of the approach. The most straightforward algorithms for generating Poisson
random variates have running time proportional to $\lambda$ \cite{knuth:vol2}, but
there exist methods whose run-time is independent of the mean \cite{gentle:03}. In
any case, as $\lambda_\mathrm{tot} = 2\beta J_\mathrm{tot}$ this is $O(E)$, in
general, but reduces to $O(N)$ for the relevant case of energy-integrable couplings
(i.e., $\sigma > 0$ for the Ising FM in one dimension). In this case, the total
number of events to distribute is $O(N)$. The distribution of events in the second
step is performed using a look-up operation. Specifically, if we use the shorthand
$p_k$, $k=1,\ldots,E$ for the probabilities $\lambda_{ij}/\lambda_\mathrm{tot}$, one
draws a random number $r$ uniformly in $[0,1[$. If $r < p_1$ the event is allocated
to edge 1. Alternatively, if $r < p_1+p_2$ the event is allocated to edge 2
etc. While this approach has $O(E)$ scaling, it can easily be sped up by a binary
(bisection) search in the table $\sum_{i<j} p_i$ of cumulative probabilities,
bringing the computational effort down to $O(\ln E)$.

As it turns out, however, also this simplification is not optimal and a faster method
is provided by Walker's method of alias that can be outlined as follows: one first
sets up a table $U_k = Ep_k$ and tries to sample from the distribution $p_k$ by
selecting one of the entries in the table with a uniform random number in
$[1,E]$. This would only be correct, however, if each $U_k = 1$. In reality, there
are ``over-full'' bins $U_k > 1$ and ``under-full'' bins $U_k < 1$. One now starts a
procedure of re-distributing the extraneous weight from over-full bins to under-full
ones, keeping track of the origin of weights using an alias table $A_k$. Once these
tables are set up, perfect samples can be drawn from $p_k$ using just two uniform
random numbers, one to index into $U_k$ and a second one for the alias table
$A_k$. Details can be found in Refs.~\cite{knuth:vol2,fukui:09}. This approach
provides look-ups and distribution of an event in constant time.

For the cluster identification it is not convenient to store the bond states as is
sometimes done for short-range models as this would bring the computational (and
storage) effort up to $O(N^2)$ again. Instead, we make use of the tree-based
union-and-find method, where the cluster structure is stored as a forest of trees
(implemented as an array of pointers). Each time a previously deleted bond is
activated, connectivity queries decide whether the connected nodes belong to
different trees, in which case one of the trees is attached to the other at the
current leaf. While for a naive implementation connectivity queries and bond
insertion have $O(N)$ scaling, additional heuristics known as path compression and
tree balancing bring the run-time scaling down to $O(\log N)$ if employing one of
these heuristics or even almost $O(1)$ if both tricks are combined\footnote{Here,
  `almost' refers to an extremely slow $N$ dependence that is derived in
  Ref.~\cite{tarjan:75}.}. Details of this approach can be found in
Refs.~\cite{knuth:vol2,newman:01a,elci:13}.

As a result, the Fukui-Todo approach shows run-time scaling that is, for all
practical purposes, indistinguishable from $O(N)$ for systems with energy convergent
couplings. This includes the mean-field model, where couplings are normally chosen to
be $J_{ij} = 1/N$ to ensure a finite energy in the thermodynamic limit. The storage
requirement is $O(E)$ for the look-up table in the bond distribution step, whereas
further storage requirements are $O(N)$ and therefore dominated by those of the
tables.  We note that for the case of translationally invariant systems we can reduce
the size of the look-up table to $O(N)$ as only the distance between sites matters,
and in the bond distribution step we can choose a site $i$ at random as well as a
distance $k$ using the alias method on the look-up table to find a partner site $j$
to increment $n_{ij}$.

\subsection{Single-cluster method}
\label{sec:FTsingle}

Single-cluster methods are typically more efficient than multi-cluster ones, and it
is indeed possible to formulate a single-cluster variant of the Fukui-Todo approach
as we will now show. The composition property (\ref{eq:composition}) of the Poisson
distribution can also be used to separately decide about how many events are to be
distributed over the bonds connecting to a specific site $i$,
\begin{equation}
  \prod_{i,j} f_{\lambda_{ij}}(n_{ij}) = \prod_i
  f_{\lambda_i}(n_i)
  \frac{n_i!}{\prod_{j\ne i}n_{ij}!}
  \prod_{j \ne i} \left(\frac{\lambda_{ij}}{\lambda_i}\right)^{n_{ij}}.
  \label{eq:composition2}
\end{equation}
Note that on the left-hand side we now consider the product over pairs $i$, $j$ of
sites, while in Eq.~(\ref{eq:composition}) the product was over bonds $(i,j)$, so in
Eq.~(\ref{eq:composition2}) each bond is counted twice. This identity is valid for
$\lambda_i = \beta J_i = \beta \sum_{j\ne i} J_{ij}$ and we have adopted the
notation $n_i = \sum_{j\ne i} n_{ij}$. It is hence possible to draw a number $n_i$ of
events for each site according to the Poisson distribution $f_{\lambda_i}(n_i)$ and
distribute them onto the bonds adjacent to site $i$ according to the probability
$\lambda_{ij}/\lambda_i$ using, e.g., the alias method. This generates exactly the
same multi-cluster dynamics as the algorithm discussed in Sec.~\ref{sec:FTmulti}.

At the same time, however, the weight decomposition~(\ref{eq:composition2}) per site
naturally suggests a single-cluster variant of the algorithm. We start from the
observation that a logically consistent definition of the single-cluster method is to
perform a full multi-cluster decomposition of the lattice and then pick a lattice
site uniformly at random and flip the cluster to which the site belongs. We now focus
on this cluster and attempt to construct it without the full multi-cluster
decomposition. We pick a seed site at random, draw the relevant number of events
$n_i$ according to $f_{\lambda_i}$ and distribute them onto the bonds $n_{ij}$
emanating from site $i$. For each $n_{ij}$ that gets an event, we put site $j$ onto a
stack of sites belonging to the cluster. We then fetch the next site from the stack
and proceed with it in the same way as with the seed site. The process terminates if
the stack is empty. While at first sight this might appear to construct a cluster
according to the generalized FK measure (\ref{eq:RCMgen}) it misses the fact that for
a site $j$ that is ultimately {\em not\/} part of the cluster according to this
construction no bond events are ever generated and distributed, thus underestimating
the probability that $j$ is added to the cluster. This is a consequence of the fact
that in this scheme each bond $(i,j)$ has {\em two\/} chances to receive events, once
when inspecting $i$ and once when inspecting $j$. We can correct for this bias by
creating {\em two\/} events for each bond emanating from a cluster site,
corresponding to $\lambda_i' = 2\lambda_i$. As a side effect, this also doubles the
average number of events on bonds that are between sites inside of the cluster, but
this does not affect the final cluster composition as all $n_{ij} > 1$ are equivalent
for the cluster identifcation. We hence arrive at the following single-cluster
algorithm:
\begin{enumerate}
\item Choose a seed site $i$ uniformly, put it onto the stack and flip $s_i$.
\item If the stack is non-empty, remove the topmost site, $i$;
  otherwise terminate the algorithm.
\item Generate an integer $n_i > 0$ randomly from the Poisson distribution
  $f_{2\lambda_i}$ and distribute $n_i$ events over the bonds $(i,j)$ with probability
  $\lambda_{ij}/\lambda_i$ using the alias method.
\item Put all such sites $j$ for which $n_{ij} > 0$ and $s_i \ne s_j$ onto the stack
  and flip $s_j$.
\item Goto step 2.
\end{enumerate}

As for the multi-cluster variant, the look-up table simplifies for the translationally
invariant system. We see that this algorithm is significantly simpler than the
multi-cluster variant as the cluster identification does not require the tree-based
union-and-find algorithm. The run-time is strictly linear for all cases where
$J_\mathrm{tot} = O(N)$, i.e., for models with convergent total energy.

\section{Dynamical scaling}
\label{sec:scaling}

We now tend to an empirical analysis of the available algorithms for the case of the
power-law model according to Eqs.~\eqref{eq:hamiltonian} and \eqref{eq:potential}. As
outlined above, the efficiency of a Markov chain Monte Carlo algorithm comprises the
two aspects of (1) the computational time required per update and (2) the scaling of
relaxation or autocorrelation times, in particular for simulations in the vicinity of
continuous phase transitions.

To study the second aspect, consider the time series $A_t$, $t=1$, $\ldots$, $N$ of
measurements of an observable $A$. In thermal equilibrium, the autocorrelation
function is time-translation invariant and exhibits an asymptotically exponential
decay,
\begin{equation}
  C_{\Delta t} = \langle A_t A_{t+\Delta t} \rangle - \langle A_t\rangle \langle
  A_{t+\Delta t} \rangle \sim e^{-\Delta t/\tau_\mathrm{exp}}.
  \label{eq:autocorr}
\end{equation}
The sampling efficiency is determined by the size of statistical fluctuations in the
final averages. For a set of $N$ {\em independent\/} measurements, we know that 
\[
\sigma^2_\mathrm{uncorr}(\bar{A}) = \frac{\sigma^2(A)}{N},
\]
but in the presence of correlations between subsequent measurements of the form
\eqref{eq:autocorr} we find instead \cite{janke:02}
\[
\sigma^2(\bar{A}) = \frac{\sigma^2(A)}{N_\mathrm{eff}},\;\;\; N_\mathrm{eff} = N/2\tau_\mathrm{int},
\]
where
\begin{equation}
  \tau_\mathrm{int} = \frac{1}{2} + \sum_{\Delta t=1}^{N-1}\frac{C_{\Delta t}}{C_0}
  \left(1-\frac{\Delta t}{N}\right)
  \label{eq:tauint}
\end{equation}
is the {\em integrated autocorrelation time\/}. In the vicinity of a critical point,
we expect dynamical scaling of autocorrelation times according to \cite{binder:book2}
\[
\tau_\mathrm{int} \sim \xi^{z_\mathrm{int}},
\]
where $\xi$ is the spatial correlation length and $z_\mathrm{int}$ denotes the
dynamical critical exponent. The value of $z_\mathrm{int}$ depends on the model under
consideration as well as the Monte Carlo algorithm. For short-range interactions and
local updates, one in general expects diffusive propagation of information, implying
a coupling of time and length scales according to $z_\mathrm{int} \approx 2$. In mean
field one can in fact show that $z_\mathrm{int} = 2$ exactly \cite{persky:96}. For
cluster algorithms for short-range models, one finds significantly reduced values
such as $z_\mathrm{int} = 0.14(1)$ and $z_\mathrm{int} = 0.46(3)$ \cite{deng:07a} for
the Swendsen-Wang algorithm for the 2D and 3D Ising models, respectively, and
$z_\mathrm{int} \approx 0.26$ (2D) and $z_\mathrm{int} \approx 0.28$ (3D) for the
Wolff algorithm \cite{wolff:89}\footnote{Note that the estimates for the Wolff update
  are significantly older than those for Swendsen-Wang and it is now believed that in
  reality $z_\mathrm{int}^\mathrm{Wolff} \le z_\mathrm{int}^\mathrm{SW}$ also in
  2D.}. For the mean-field model, rigorous arguments imply that
$z_\mathrm{int}^\mathrm{SW} = 1$ and $z_\mathrm{int}^\mathrm{Wolff} = 0$
\cite{ray:89,persky:96}.

\begin{figure}
  \begin{center}
    \includegraphics[scale=0.8]{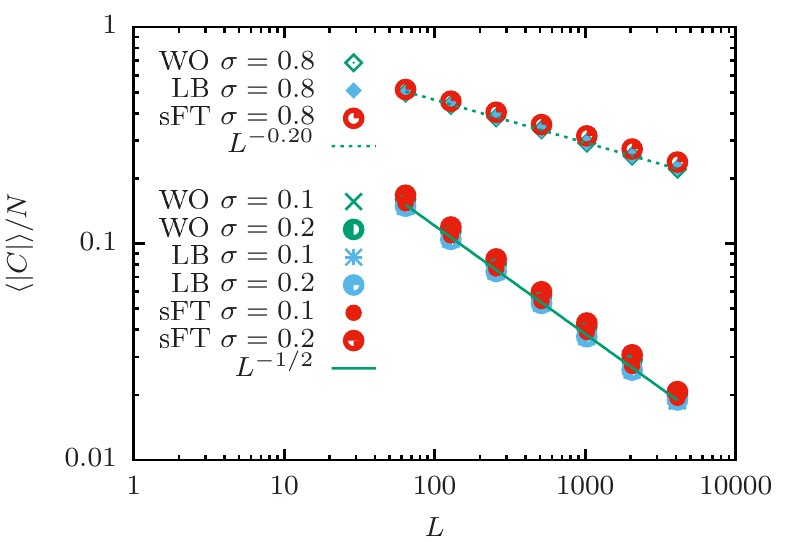}
    \includegraphics[scale=0.8]{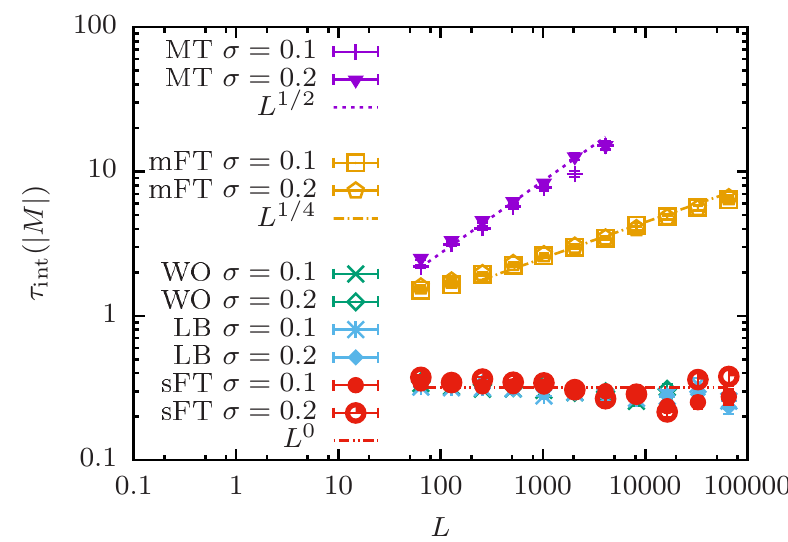}
  \end{center}
  \caption{\label{fig:cluster-size}%
    Left: Average relative size $\langle |C|\rangle/N$ of clusters grown in the
    single-cluster algorithms of the plain Wolff type (WO), the Luijten-Bl\"ote
    algorithm (LB) and the single-cluster Fukui-Todo update (sFT) for simulations of
    the 1D long-range critical Ising model at the critical temperature in the
    non-mean-field ($\sigma = 0.8$) and mean-field ($\sigma = 0.1$ and
    $\sigma = 0.2$) region of the interaction range. Right: Integrated
    autocorrelation times of the magnetization for $\sigma = 0.1$ and $\sigma = 0.2$
    in the mean-field regime below the critical value $\sigma = 1/2$ for the
    Metropolis (MT), multi-cluster Fukui-Todo (mFT), Wolff (WO), Luijten-Bl\"ote (LB)
    and single-cluster Fukui-Todo (sFT) updates as a function of system size $L$. }
\end{figure}

For the model with power-law interactions, a study of the Langevin dynamics for the
spherical model yields $z_\mathrm{int}^\mathrm{Metro} = \sigma$
\cite{cannas:01}. This carries over to the Ising model in the mean-field regime
$0 < \sigma < 1/2$. By analogy to the mean-field limit, we conjecture that
$z_\mathrm{int}^\mathrm{SW} = \sigma/2$ for multi-cluster and
$z_\mathrm{int}^\mathrm{Wolff} = 0$ for single-cluster updates in the same
regime. Due to the particular scaling of the correlation length with system size
according to $\xi \sim L^\qqq$ above the upper critical dimension \cite{berche:12a},
where $\qq = d/d_c$ and $d_c = 4$ for short-range models and $\qq = d/2\sigma$ for
the interactions \eqref{eq:potential} \cite{flores:15}, we expect the following
finite-size scaling of the autocorrelation times,
\begin{equation}
  \tau_\mathrm{int} \sim L^{z_\mathrm{int}\qqq} \sim
  \left\{
    \begin{array}{l@{\hspace{0.5cm}}l}
      L^{1/2} & \mbox{for Metropolis}, \\
      L^{1/4} & \mbox{for multi cluster}, \\
      L^0 & \mbox{for single cluster}.
    \end{array}
  \right.
  \label{eq:dynamical-scaling}
\end{equation}
We note that this is consistent with the behavior in the limit $\sigma \to 0$, where
the model becomes equivalent to a mean-field system, and a scaling
$\tau_\mathrm{int} \sim L^{z_\mathrm{int}/4}$ is observed according to the usual
identification of $N^{1/4}$ with the linear system size in the mean-field case
\cite{persky:96}.

We have determined the integrated autocorrelation times of the energy and
magnetization for the 1D power-law model using a standard self-consistent cut-off
procedure for the summation of the autocorrelation function given in
Eq.~\eqref{eq:tauint}. For a discussion of this procedure, including the estimation
of statistical errors see, e.g., Ref.~\cite{janke:02}. Our simulations were performed
for systems with periodic boundary conditions and for the interaction-range exponents
$\sigma = 0.1$ and $\sigma = 0.2$ in the mean-field range $\sigma < 1/2$ as well as a
number of values $\sigma \ge 0.5$ in the non-trivial long-range regime. We used an
Ewald summed form of the interaction \eqref{eq:potential} \cite{flores:15} and
performed simulations at the previously determined critical temperatures
$T_c=21.0013$ for $\sigma = 0.1$, $T_c=10.8419$ for $\sigma=0.2$, $T_c=4.36395$ for
$\sigma = 0.5$, $T_c=3.54886$ for $\sigma = 0.6$, $T_c=2.93061$ for $\sigma = 0.7$,
$T_c=2.43267$ for $\sigma = 0.8$, and $T_c=2.00144$ for $\sigma = 0.9$
\cite{luijten:97b}. To provide autocorrelation times on a common time scale for all
algorithms it is convenient to consider updates such that, on average, each spin is
touched once (one sweep). While this is automatic for the Metropolis and
multi-cluster updates, for the single-cluster variants it implies a rescaling of the
raw autocorrelation times $\tau_\mathrm{int}'$ determined from a time series recorded
after each individual single-cluster update according to
\[
\tau_\mathrm{int}^\mathrm{sc} = \tau_\mathrm{int}' \frac{\langle |C|\rangle}{N},
\]
where $\langle|C|\rangle$ denotes the average size of the simulated clusters. It is
known that $\langle|C|\rangle$ provides an improved estimator of the susceptibility
$\chi$ \cite{wj:chem}. We hence expect a scaling of
$\langle |C|\rangle \sim \xi^{\gamma/\nu}$ and thus
$\langle |C|\rangle/N \sim L^{\qqq\gamma/\nu-d}$. For the mean-field cases
$\sigma = 0.1$ and $\sigma = 0.2$ the values $\gamma/\nu = \sigma$ \cite{flores:15}
and $\qq = 1/2\sigma$ imply $\qqq\gamma/\nu-d = -1/2$. What is more, the long-range
model has the pecularity that the value of $\gamma/\nu = 2-\eta = \sigma$ does not
acquire any corrections beyond mean-field even for $1/2 \le \sigma \le 1$, such that
$\langle |C|\rangle/N$ scales as $L^{\sigma-1}$ there. These theoretical
considerations are fully confirmed by the simulation results for the average cluster
size summarized in Fig.~\ref{fig:cluster-size} (left panel).

\begin{figure}
  \begin{center}
    \includegraphics[scale=0.8]{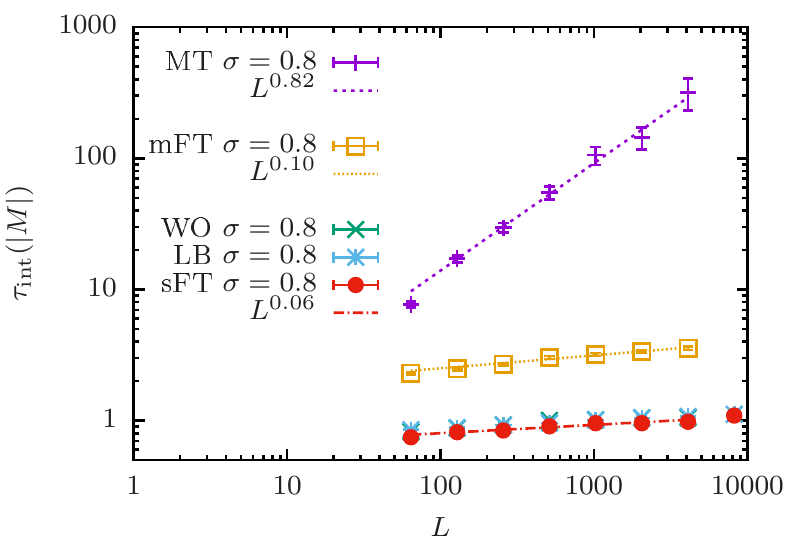}
    \includegraphics[scale=0.8]{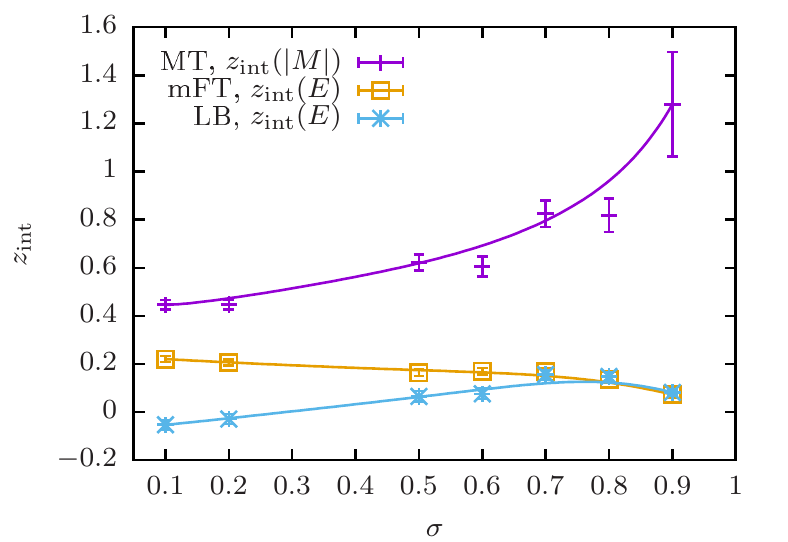}
  \end{center}
  \caption{\label{fig:nonMF}%
    Left: Integrated autocorrelation times for the magnetization for the Metropolis
    (MT), multi-cluster Fukui-Todo (mFT), Wolff (WO), Luijten-Bl\"ote (LB) and
    single-cluster Fukui-Todo (sFT) updates at $\sigma = 0.8$ in the non-trivial
    long-range regime. Right: Dynamical critical exponents $z_\mathrm{int}$ of the
    magnetization for the Metropolis update (MT) and of the internal energy for the
    multi-cluster Fukui-Todo (mFT) and Luijten-Bl\"ote (LB) updates as a function of
    interaction range $\sigma$. The lines are Bezier curves to guide the eye.
  }
\end{figure}

Taking this rescaling into account for the single-cluster variants, the right panel
of Fig.~\ref{fig:cluster-size} shows our numerical results for the integrated
autocorrelation times per sweep of the (modulus of the) magnetization for the values
$\sigma = 0.1$ and $\sigma = 0.2$ in the mean-field regime $0\le \sigma \le 1/2$. We
find excellent agreement with the theoretical prediction \eqref{eq:dynamical-scaling}
with some signs of the presence of scaling corrections for smaller system sizes. In
particular, the results confirm that the Wolff algorithm (WO), the Luijten-Bl\"ote
algorithm (LB) and the single-cluster variant of the Fukui-Todo update (sFT)
introduced here exhibit the same asymptotic dynamical behavior which is substantially
superior to the dynamical behavior of the multi-cluster approach (mFT) and, even more
so, the Metropolis update (MT). While the data in Fig.~\ref{fig:cluster-size} show
the autocorrelation times of the magnetization, we have also determined those of the
internal energy and find practically identical results there (not shown).

Performing similar runs for the non-trivial long-range value $\sigma = 0.8$, we
arrive at different estimates of $z_\mathrm{int}$, cf.\ the data shown in the left
panel of Fig.~\ref{fig:nonMF}. In this regime we also start to see some differences
in the scaling of autocorrelation times for internal energies and magnetizations. It
can be shown that for single-spin flip heatbath dynamics the magnetization
corresponds to the slowest mode, while for a random-cluster single-bond update
(Sweeny's algorithm \cite{sweeny:83,elci:13}) the slowest mode is given by the bond
density corresponding to the internal energy \cite{elci:15b}. We expect the same to
be true for the single-spin flip Metropolis update and the various cluster algorithms
considered here. We indeed find numerically that our estimates
$z_\mathrm{int}(|M|) > z_\mathrm{int}(E)$ for Metropolis, while in contrast
$z_\mathrm{int}(|M|) < z_\mathrm{int}(E)$ for the cluster updates in the regime
$\sigma \ge 0.5$. The right panel of Fig.~\ref{fig:nonMF} summarizes our results for
$z_\mathrm{int}(|M|)$ for Metropolis and $z_\mathrm{int}(E)$ for multi-cluster and
single-cluster dynamics. Note that the single-cluster algorithms WO, LB and sFT
implement the same cluster dynamics, so result in the same estimates of
$z_\mathrm{int}$, and for clarity we only show one data set. We see that for
Metropolis the dynamical critical exponent is moving towards
$z_\mathrm{int} \gtrsim 2$ expected for short-range models, while the behavior of
multi-cluster and single-cluster updates appears to coalesce at a value of
$z_\mathrm{int}$ close to $0$. We did not perform simulations at the upper critical
interaction range $\sigma = 1$ as there the system undergoes a Kosterlitz-Thouless
phase transition \cite{angelini:14}, and we expect strong scaling corrections. For
$\sigma > 1$ there is no finite-temperature phase transition \cite{dyson:69}.

\begin{figure}
  \begin{center}
    \includegraphics[scale=0.8]{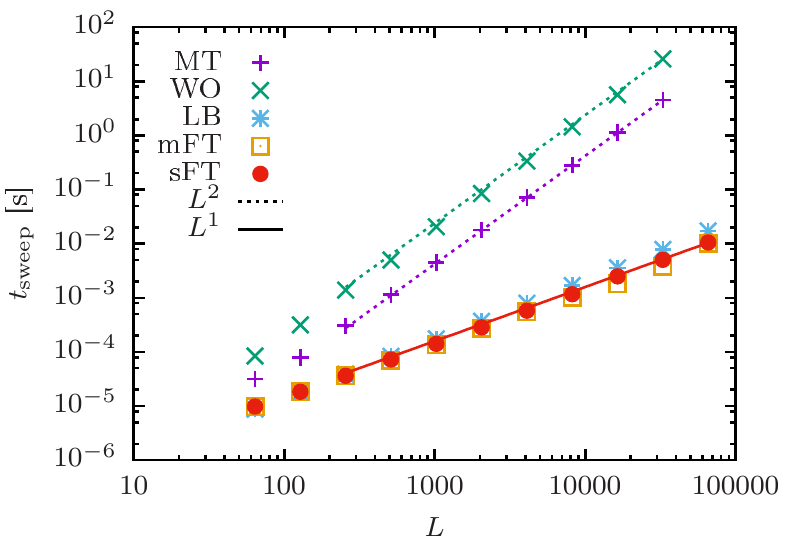}
    \includegraphics[scale=0.8]{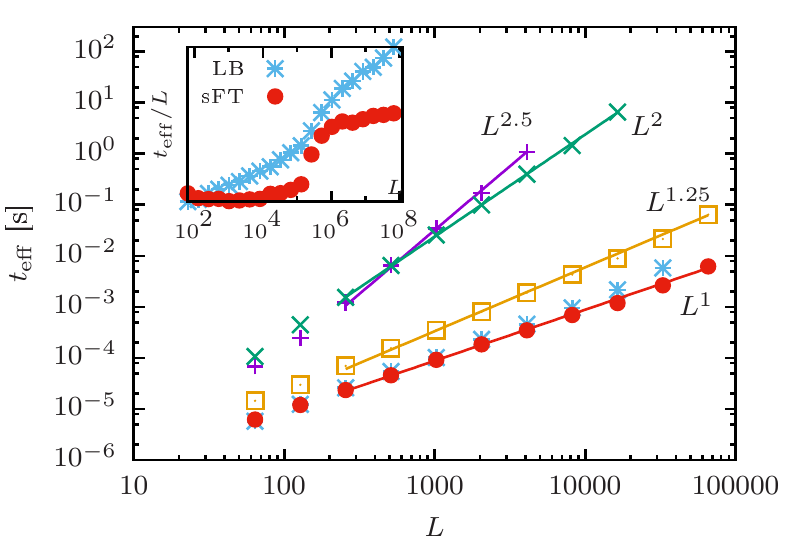}
  \end{center}
  \caption{\label{fig:times}%
    Left: Times in seconds per sweep on an Intel Core i5 3210M CPU for the different
    update algorithms applied to the 1D long-range Ising model with $\sigma = 0.1$ at
    criticality and fits of the quadratic and linear laws to the data. Right: Time
    $t_\mathrm{eff}$ to generate an effective independent sample of the model with
    different update algorithms. The inset shows the normalized time
    $t_\mathrm{eff}/L$ per spin comparing the two most efficient approaches, LB and
    sFT. The legend is the same as that of the plot on the left.}
\end{figure}

We finally consider the behavior of the actual run-time per update for the different
algorithms. While such times are clearly hardware specific, the scaling of the
different algorithms is not, and so this is the aspect we focus on here. The left
panel of Fig.~\ref{fig:times} summarizes the run-times per sweep for the different
algorithms run on the same hardware, an Intel Core i5 3210M CPU running at
2.50GHz. The asymptotically quadratic run-times of the MT and WO algorithms are
clearly visible. For the LB approach a slightly steeper than linear increase of
run-times is seen, but as expected it is difficult to resolve the additional
logarithmic component explicitly. The behavior of the mFT approach is compatible with
linear scaling in the range of system sizes considered. Finally, the sFT algorithm
follows the expected linear scaling in $L$. For the overall efficiency of the
considered updates, it is the combination of computational effort per sweep and the
achieved integrated autocorrelation times that matters. We hence consider the
quantity
\[
t_\mathrm{eff} = t_\mathrm{sweep}\tau_\mathrm{int},
\]
which is the wall-clock time required to generate a statistically independent sample,
as the final measure of efficiency \cite{elci:13}. These times are shown in the right
panel of Fig.~\ref{fig:times} for the algorithms considered here. As the lines in the
plot show, the expected scaling from the computational complexity of the algorithms
and the dynamical behavior according to Eq.~\eqref{eq:dynamical-scaling} of
$\sim L^{2.5}$ for Metropolis, $\sim L^2$ for Wolff, $\sim L^{1.25}$ for
multi-cluster Fukui-Todo and $\sim L^1$ for the single-cluster Fukui-Todo approaches
is fully compatible with the numerical data. It is clearly visible that the LB
algorithm and the sFT approach introduced here show the asymptotically best
performance, with an advantage over the other approaches that grows algebraically
with $L$. As the inset displaying the time $t_\mathrm{eff}/L$ per spin shows, the sFT
approach has perfect linear algorithmic scaling , while the LB algorithm has a
logarithmic overhead for the weight look-up. The pronounced step in the run-times of
both algorithms is due the fact that at a given, hardware-dependent system size the
look-up table starts to exceed the size of the cache memory. At the largest system
size considered here, $L = 2^{26} \approx 7\times 10^7$, the sFT approach is about
three times faster than LB. Comparing to the other algorithms, we note that already
for $L=65\,536$ the sFT algorithm is about a million times faster to produce an
independent sample than the Metropolis method.

\section{Summary and outlook}
\label{sec:conclusions}

We have discussed a range of different algorithms for the simulation of spin systems
with long-range interactions. Naive approaches exhibit unfavorable $O(N^2)$ scaling
with the number of spins $N$, but it is possible to formulate cluster-update
algorithms that combine $O(N\ln N)$ or even $O(N)$ scaling of the run-time per sweep
with the additional benefit of a reduced critical slowing down of systems close to
continuous phase transitions. The scaling of autocorrelation times in the mean-field
regime $0\le \sigma \le 1/2$ of the 1D power-law Ising model is explained in terms of
the modified QFSS approach to finite-size scaling \cite{berche:12a,flores:15}. We
introduced a single-cluster algorithm based on the generalized Fortuin-Kasteleyn
representation \eqref{eq:RCMgen} that is the only known algorithm with strictly
linear scaling of run times. It outperforms all previously known approaches and, at
the same time, is very straightforward to implement and can be applied to systems in
any dimension with or without translational invariance. In the non-trivial long-range
regime $1/2 \le \sigma \le 1$, we observed a continuous variation of dynamical
critical exponents with the Metropolis exponent wandering in the direction of the
established value $z \gtrsim 2$ expected for short-range models and the values for
single-cluster and multi-cluster dynamics coming closer together as the upper
critical range $\sigma = 1$ is approached.

Some important aspects of the problem have been omitted from the present
discussion. This includes the fact that for such systems with periodic boundary
conditions a summation over an infinite number of interaction partners is necessary
to get reliable results. The corresponding Ewald summation is discussed, e.g., in
Refs.~\cite{flores:15,emilio:thesis}. Another aspect is the problem of measuring the
energy for the long-range interactions, a task that in itself has $O(N^2)$ scaling in
the straightforward approach. As was shown in Ref.~\cite{fukui:09}, a linear
algorithm can be formulated in the framework of the generalized Fortuin-Kasteleyn
representation \eqref{eq:RCMgen}. The algorithms discussed here can also be
generalized for the presence of external magnetic fields \cite{emilio:thesis}, but
they become less efficient with increasing field strength. While a substantial
reduction of critical slowing down from cluster updates can only be expected for
non-frustrated systems, the reduction of computational complexity of the present
approach as compared to local updates might also make it interesting for simulations
of spin-glass systems with long-range interactions \cite{beyer:12}.

\section*{Author contribution statement}

EFS and MW performed the simulations, carried out the analysis and wrote the paper.
The paper emerged out of joint work in collaboration with BB and RK.

\section*{Acknowledgments}

The article is dedicated to Wolfhard Janke on the occasion of his 60th
birthday. M.W. acknowledges some useful discussions with Synge Todo. The
implementation of numerical calculations was personally encouraged by Jim Tabor. This
work was supported by the EU FP7 IRSES network DIONICOS under contract No.\
PIRSES-GA-2013-612707 and by the Coll\`ege Doctoral ``Statistical Physics of Complex
Systems'' Leipzig-Lorraine-Lviv-Coventry (${\mathbb L}^4$).


\begin{thebibliography}{10}

\bibitem{kawashima:03a}
N. Kawashima and H. Rieger,  in {\em Frustrated Spin Systems}, edited by H.~T.
  Diep (World Scientific, Singapore, 2005), Chap.~9, p.\ 491.

\bibitem{fytas:16}
N.~G. Fytas, V. Mart\'{\i}n-Mayor, M. Picco, and N. Sourlas, Phys. Rev. Lett.
  {\bf 116},  227201  (2016).

\bibitem{wittmann:14}
M. Wittmann and A.~P. Young, Phys. Rev. E {\bf 90},  062137  (2014).

\bibitem{flores:16}
E. Flores-Sola, B. Berche, R. Kenna, and M. Weigel, Phys. Rev. Lett. {\bf 116},
   115701  (2016).

\bibitem{fisher:74a}
M.~E. Fisher, Rev. Mod. Phys. {\bf 46},  597  (1974).

\bibitem{rapaport:04}
D.~C. Rapaport, {\em The Art of Molecular Dynamics Simulation} (Cambridge
  University Press, Cambridge, 2004).

\bibitem{binder:book2}
K. Binder and D.~P. Landau, {\em A Guide to Monte Carlo Simulations in
  Statistical Physics}, 4th ed. (Cambridge University Press, Cambridge, 2015).

\bibitem{metropolis:53a}
N. Metropolis {\it et~al.}, J. Chem. Phys. {\bf 21},  1087  (1953).

\bibitem{swendsen-wang:87a}
R.~H. Swendsen and J.~S. Wang, Phys. Rev. Lett. {\bf 58},  86  (1987).

\bibitem{wolff:89a}
U. Wolff, Phys. Rev. Lett. {\bf 62},  361  (1989).

\bibitem{luijten:06}
E. Luijten, Lect. Notes Phys. {\bf 703},  13  (2006).

\bibitem{deng:07}
Y. Deng, T.~M. Garoni, and A.~D. Sokal, Phys. Rev. Lett. {\bf 98},  230602
  (2007).

\bibitem{elci:13}
E.~M. El\c{c}i and M. Weigel, Phys. Rev. E {\bf 88},  033303  (2013).

\bibitem{lacroix:11}
C. Lacroix, P. Mendels, and F. Mila, {\em Introduction to Frustrated Magnetism:
  Materials, Experiments, Theory} (Springer, Berlin, 2011), Vol.~164.

\bibitem{britton:12}
J.~W. Britton {\it et~al.}, Nature {\bf 484},  489  (2012).

\bibitem{fisher:72}
M.~E. Fisher, S.~K. Ma, and B.~G. Nickel, Phys. Rev. Lett. {\bf 29},  917
  (1972).

\bibitem{dyson:69}
F.~J. Dyson, Commun. Math. Phys. {\bf 12},  91  (1969).

\bibitem{dyson:71}
F.~J. Dyson, Commun. Math. Phys. {\bf 21},  269  (1971).

\bibitem{xu:93}
H.-J. Xu, B. Bergersen, and Z. R{\'{a}}cz, Phys. Rev. E {\bf 47},  1520
  (1993).

\bibitem{nagle:70}
J.~F. Nagle and J.~C. Bonner, J. Phys. C {\bf 3},  352  (1970).

\bibitem{glumac:89}
Z. Glumac and K. Uzelac, J. Phys. A {\bf 22},  4439  (1989).

\bibitem{luijten:95}
E. Luijten and H.~W. Bl\"{o}te, Int. J. Mod. Phys. C {\bf 6},  359  (1995).

\bibitem{fukui:09}
K. Fukui and S. Todo, J. Comp. Phys. {\bf 228},  2629  (2009).

\bibitem{wj:chem}
W. Janke,  in {\em Computational Physics}, edited by K.~H. Hoffmann and M.
  Schreiber (Springer, Berlin, 1996), pp.\ 10--43.

\bibitem{fortuin:72a}
C.~M. Fortuin and P.~W. Kasteleyn, Physica {\bf 57},  536  (1972).

\bibitem{edwards:88a}
R.~G. Edwards and A.~D. Sokal, Phys. Rev. D {\bf 38},  2009  (1988).

\bibitem{weigel:10b}
M. Weigel, Phys. Rev. E {\bf 84},  036709  (2011).

\bibitem{luijten:97}
E. Luijten and H.~W.~J. Bl\"{o}te, Phys. Rev. B {\bf 56},  8945  (1997).

\bibitem{knuth:vol2}
D.~E. Knuth, {\em The Art of Computer Programming, Volume 2: Seminumerical
  Algorithms}, 3rd ed. (Addison-Wesley, Upper Saddle River, NJ, 1997).

\bibitem{gentle:03}
J.~E. Gentle, {\em Random number generation and Monte Carlo methods}, 2nd ed.
  (Springer, Berlin, 2003).

\bibitem{tarjan:75}
R.~E. Tarjan, J. ACM {\bf 22},  215  (1975).

\bibitem{newman:01a}
M.~E.~J. Newman and R.~M. Ziff, Phys. Rev. E {\bf 64},  016706  (2001).

\bibitem{janke:02}
W. Janke,  in {\em Proceedings of the Euro Winter School ``Quantum Simulations
  of Complex Many-Body Systems: From Theory to Algorithms''}, Vol.~10 of {\em
  NIC Series}, edited by J. Grotendorst, D. Marx, and A. Muramatsu (John von
  Neumann Institute for Computing, J\"{u}lich, 2002), pp.\ 423--445.

\bibitem{persky:96}
N. Persky, R. Ben-Av, I. Kanter, and E. Domany, Phys. Rev. E {\bf 54},  2351
  (1996).

\bibitem{deng:07a}
Y. Deng {\it et~al.}, Phys. Rev. Lett. {\bf 99},  055701  (2007).

\bibitem{wolff:89}
U. Wolff, Phys. Lett. B {\bf 228},  379  (1989).

\bibitem{ray:89}
T.~S. Ray, P. Tamayo, and W. Klein, Phys. Rev. A {\bf 39},  5949  (1989).

\bibitem{cannas:01}
S.~A. Cannas, D.~A. Stariolo, and F.~A. Tamarit, Physica A {\bf 294},  362
  (2001).

\bibitem{berche:12a}
B. Berche, R. Kenna, and J.~C. Walter, Nucl. Phys. B {\bf 865},  115  (2012).

\bibitem{flores:15}
E.~J. Flores-Sola, B. Berche, R. Kenna, and M. Weigel, Eur. Phys. J. B {\bf
  88},  1  (2015).

\bibitem{luijten:97b}
E. Luijten, Ph.D. thesis, Delft University of Technology, 1997.

\bibitem{sweeny:83}
M. Sweeny, Phys. Rev. B {\bf 27},  4445  (1983).

\bibitem{elci:15b}
E.~M. El\c{c}i, Ph.D. thesis, Coventry University, 2015.

\bibitem{angelini:14}
M.~C. Angelini, G. Parisi, and F. Ricci-Tersenghi, Phys. Rev. E {\bf 89},
  062120  (2014).

\bibitem{emilio:thesis}
E. Flores-Sola, Ph.D. thesis, Coventry University, Coventry, 2016.

\bibitem{beyer:12}
F. Beyer, M. Weigel, and M.~A. Moore, Phys. Rev. B {\bf 86},  014431  (2012).

\end{thebibliography}

\end{document}